\documentclass[]{aa} 
\usepackage[varg]{txfonts}
\usepackage[colorlinks=true, urlcolor=blue,citecolor=blue,linkcolor=blue]{hyperref}
\usepackage{graphicx}
\usepackage{natbib}
\usepackage{subcaption}
\usepackage{supertabular}
\usepackage{multirow}
\usepackage{wasysym}
\usepackage{soul}
\usepackage{footnote}
\usepackage{fixmath}
\def\kms{\mbox{km s$^{-1}$}}
\def\ms{\mbox{m s$^{-1}$}}
\def\halpha{{H$\alpha$}}

\def\HI{\ion{H}{I}}
\def\FeI{\ion{Fe}{I}}
\def\CaIIH{\ion{Ca}{II}~H}
\def\CaIIK{\ion{Ca}{II}~K}
\def\CaIIHK{\ion{Ca}{II}~H\&K}
\def\CaII{\ion{Ca}{II}}

\newcommand{\bea}{\begin{eqnarray}}
\newcommand{\eea}{\end{eqnarray}}
\newcommand{\grass}{}

\newcommand{\secgrass}{}

\usepackage{natbib,twoopt}
\bibpunct{(}{)}{;}{a}{}{,} %% natbib format for A&A and ApJ
\makeatletter
\newcommandtwoopt{\citeads}[3][][]{\href{http://adsabs.harvard.edu/abs/#3}%
{\def\hyper@linkstart##1##2{}%
\let\hyper@linkend\@empty\citealp[#1][#2]{#3}}}
\newcommandtwoopt{\citepads}[3][][]{\href{http://adsabs.harvard.edu/abs/#3}%
{\def\hyper@linkstart##1##2{}%
\let\hyper@linkend\@empty\citep[#1][#2]{#3}}}
\newcommandtwoopt{\citetads}[3][][]{\href{http://adsabs.harvard.edu/abs/#3}%
{\def\hyper@linkstart##1##2{}%
\let\hyper@linkend\@empty\citet[#1][#2]{#3}}}
\newcommandtwoopt{\citeyearads}[3][][]%
{\href{http://adsabs.harvard.edu/abs/#3}
{\def\hyper@linkstart##1##2{}%
\let\hyper@linkend\@empty\citeyear[#1][#2]{#3}}}
\makeatother

\begin{document}

\title{Chromospheric observations and magnetic configuration of a supergranular structure
%The magnetic configuration above a supergranular structure
%Observations of an unusual magnetic structure above a supergranule
%Dynamic chromosphere above a supergranular structure
}
\titlerunning{Chromospheric observations and magnetic configuration of a supergranular structure
}
\authorrunning{C. Robustini et al.}

\author{Carolina Robustini\inst{1}, Sara Esteban Pozuelo\inst{1}, Jorrit Leenaarts\inst{1}, Jaime de la Cruz Rodr{\'i}guez\inst{1}}

\institute{Institute for Solar Physics, Department of Astronomy,
  Stockholm University,
AlbaNova University Centre, SE-106 91 Stockholm Sweden \email{carolina.robustini@astro.su.se}}

\date{Received; accepted}
\abstract

\abstract {Unipolar magnetic regions are often associated with supergranular cells. The chromosphere above these regions is regulated by the magnetic field, but the field structure is poorly known. \grass{In unipolar regions, the fibrillar arrangement does not always coincide with magnetic field lines, and polarimetric observations are needed to establish the chromospheric magnetic topology.}
}
   {In an active region close to the limb, we observed a unipolar annular network of supergranular size. This supergranular structure harbours a radial distribution of the fibrils converging towards \grass{its centre. We aim to improve the description of this structure by determining the magnetic field configuration and the line-of-sight velocity distribution in both the photosphere and the chromosphere.}
 }
   {We observed the supergranular structure at different heights by taking data in the \FeI~6301-6302~\AA, \halpha, \CaII~8542~\AA\ and the \CaIIHK\ spectral lines with the CRISP and CHROMIS instruments at the Swedish 1-m Solar Telescope. \grass{We performed Milne-Eddington inversions of the spectropolarimetric data of \FeI~6301-6302~\AA\ and applied the weak field approximation to \CaII~8542~\AA\ data to retrieve the magnetic field in the photosphere and chromosphere. }
We used  photospheric magnetograms of CRISP, HINODE/SP and HMI to calculate the magnetic flux. We investigated the velocity distribution using the line-of-sight velocities computed from the Milne-Eddington inversion and from Doppler shift of the K$_3$ feature in the \CaIIK\ spectral line. To describe the typical spectral profiles characterising the chromosphere above the inner region of the supergranular structure, we performed a K-mean clustering of the spectra in \CaIIK.
   }
   {\secgrass{The photospheric magnetic flux shows that the supergranular boundary has an excess of positive polarity and the whole structure is not balanced. 
   The magnetic field vector at chromospheric heights, retrieved by the weak field approximation, indicates that the field lines within the supegranular cell tend to point inwards, and might form a canopy above the unipolar region. 
   In the centre of the supergranular cell hosting the unipolar region, we observe a persistent chromospheric brightening coinciding with a strong gradient in the line-of-sight velocity.}}
   {} 
 
\keywords{Sun: magnetic field--- Sun: chromosphere --- Sun: photosphere --- methods:observational }

\maketitle
%******************************************************************************************
\section{Introduction}
%******************************************************************************************
%iniziare con pistolotto su supergranulation che nn si sa molto in generale
Supergranulation is a poorly understood phenomenon of the solar atmosphere. Its origin is still under debate and several different models have been proposed.

%concentrarsi su quello che si sa sul rapporto con magnetic network (citare il francese) e gabriel
How the magnetic field and supergranulation are dynamically related remains unclear.
However, it is well known that the velocity field in the supergranulation advects the vertical magnetic field to the boundaries of the supergranular cells where it creates a magnetic network \citepads{2010A&A...519A..58T, 2011ApJ...735...74I,2012ApJ...758L..38O,2017A&A...599A..69R}.  

One of the first models of the magnetic field above supergranular cells in quiet Sun (QS) was proposed by \citetads{1976RSPTA.281..339G}. Assuming a potential field in a unipolar cell, he  suggested that all the flux at the photosphere is concentrated on the supergranular boundaries while the inner cell is field-free. According to his model, the boundary field expands into the higher layers in a canopy shape and it becomes uniform above the cell when it reaches the corona. 
The study of \citetads{1983SoPh...87...37J} provided the first observational support to this model. They reported on six unipolar regions where the main polarity field lines fan out towards the upper layers of the atmosphere. The extent of the unipolar regions usually 
coincides with supergranule boundaries.

Further studies supported the presence of a low-lying magnetic canopy \citepads[e.g.][]{1990A&A...234..519S}. However, there is no consensus about the presence of a field canopy above supergranular cells.

Alternatively, \citetads{2003ApJ...597L.165S} proposed that a significant contribution to the photospheric flux is located in the inner supergranular cell. In such a scenario, a relatively strong internetwork can produce a  low-lying canopy of field lines connecting the mixed polarities in the internetwork to the network boundaries, and this may translate into a low-lying canopy of fibrils \citepads[see also][]{2006ApJ...647L.183A, 2006ASPC..354..276R}.

The arrangement of fibrils in the chromosphere is expected to mainly follow the magnetic field lines \citepads{2011A&A...527L...8D, 2013ApJ...768..111S, 2015ApJ...802..136L,2016ApJ...831L...1M,2016ApJ...826...51Z, 2017A&A...599A.133A}, but this dependency becomes rather unclear in unipolar regions. 
% fare un riassuntino di quello che dice reardon
\citetads{2011ApJ...742..119R} observed a unipolar region of $\sim40\arcsec$ diameter in the line core of \CaII~8542~\AA, where all the fibrils crossing the unipolar region were completely confined in the cell interior, as also described in \citetads{1971SoPh...19...59F}.
%dire in particolare che lui osserva un caso in cui sono pettinate da una parte all altra
\grass{These fibrils appear to connect same polarity concentrations as they originate from the stronger field patches in the cell boundaries.}
However \citetads{2011ApJ...742..119R} pointed out that usually only one footpoint is clearly rooted in the network, suggesting a rather complex magnetic topology above the unipolar region.

The parallel fibril arrangement observed by \citetads{2011ApJ...742..119R} is commonly known as chain, and it is one of the possible organised distributions in which fibrils are clustered into \citepads{1993A&A...271..574T}. 
An alternative distribution is called rosette and it consists of a radial arrangement of fibrils at the vertices of the supergranular cells. At this location, the magnetic field is usually quite concentrated and can reach up to a few kG. The strong magnetic field is thought to be the cause of the brightening observed in the rosette cores \citepads{1983A&A...125..280D}. \citetads{1993A&A...271..574T} measured a blue-shift in the \halpha\ spectral line core, at the very  centre  of the rosette and  on the top of the radial fibril. They interpreted this as an upflow, while they also observed a downflow located at the fibril footpoints.
Typical value for the rosette diameter ranges from $\sim10\arcsec$ \citepads{2017A&A...607A..46M} up to $\sim50\arcsec$ \citepads{2003A&A...402..361T}.

In this paper, we present high-resolution chromospheric observations of a unipolar region enclosed in a supergranular structure (SGS), with a radial arrangement of the fibrils. 
\grass{The novelty of these data lies in (1) narrowband images in three different chromospheric spectral lines, including \CaIIHK\ spectrally resolved observations with the new CHROMospheric Imaging Spectrometer (CHROMIS) at the Swedish 1-m Solar Telescope \citepads[SST,][]{2003SPIE.4853..341S}, (2) and the peculiar morphology of the observed structure. The latter} reminds of a chromospheric rosette, although the convergence point of the fibrils is located at the centre of the supergranular cell. 
\grass{Since the fibril distribution inside the unipolar cell cannot always be trusted as a magnetic field tracer, we aim to establish the magnetic field topology using polarimetry. We also aim to investigate the dynamics of the fibrils using line-of-sight (LOS) velocity information.}

%******************************************************************************************
\section{Observations and data reduction}
%******************************************************************************************
On 2017 April 20, the observed unipolar region was located close to the active region NOAA~12651 on the east side of the limb (X=-845$\arcsec$, Y=261$\arcsec$). This target was observed between 09:41 and 10:01 UT with a heliocentric angle $\approx 68^{\circ}$ ($\mu$=0.37).
Observations were carried out using the CRisp Imaging SpectroPolarimeter 
\citepads[CRISP,][]{2006A&A...447.1111S, 2008ApJ...689L..69S} 
and CHROMIS, at the SST.
CRISP data were recorded along the following wavelength positions:
\begin{itemize}
\item[•] \HI~6563 (\halpha), at 25 equidistant positions between 6560.6 and 6565.4~\AA\ with $\Delta \lambda=$ 200~m\AA\ ;
\item[•]\FeI~6301, at 10 non-equidistant positions between 6300.7 and 6301.2~\AA\  with  $\Delta \lambda=$ 40~m\AA\ except $\Delta \lambda=$ 100~m\AA\ between the three positions  in the blue wing and two positions in the red wing closest to the continuum;
\item[•]\FeI~6302, at 11 equidistant positions between between 6301.8 and 6302.2 with  $\Delta \lambda=$ 40~m\AA\
\item[•]\CaII~8542, at 17 positions between 8541.0 and 8543.0~\AA\ with $\Delta \lambda=$ 100~m\AA\ except $\Delta \lambda=$ 200~m\AA\ between the two positions in the red and in the blue wings closer to the continuum.
\end{itemize} 

The time cadence is $\sim$39~s. The pixel size and the spectral resolution $R$ at 6302~\AA\ are 0\farcs059 and $R=\lambda / \delta\lambda \approx 1.14\times10^{5}$, respectively. 
\FeI~6301-6302~\AA\ and \CaII~8542~\AA\ were recorded in full Stokes polarimetry. 
Although the seeing-induced cross-talk is corrected using two liquid crystal modulators and a beam splitter for the temporal and spatial modulation respectively, a small remaining cross-talk is still present in the Stokes profiles. Since we assumed a Zeeman regime, we have corrected the residual cross-talk by minimising the polarisation signal at the wavelength positions closest to the continuum.

\begin{figure}
\includegraphics[scale=1]{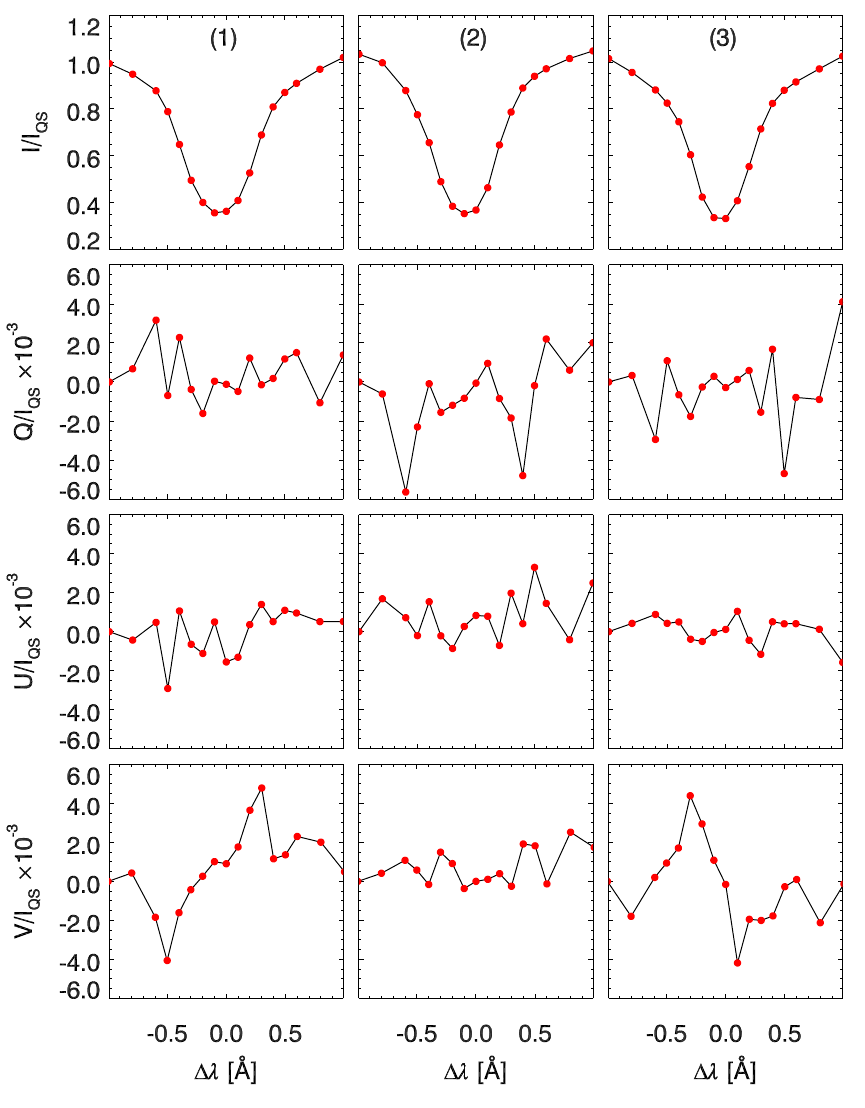}\caption{\secgrass{Typical Stokes profiles of \CaII~8542~\AA\ observed within the SGS and corresponding to pixels 1, 2, 3 in Figure~\ref{fig:magneto}b. The red dots indicate the observed wavelengths.}}\label{fig:ex_prof}
\end{figure}

\grass{In the magnetic field calculations shown in Section~\ref{sec:res}, we used the frame of the \FeI~6301-6302~\AA\ time series recorded with the best atmospheric conditions. The polarisation signal in 
\CaII~8542~\AA\ is usually weaker. To increase the signal-to-noise ratio, we computed the median of nine frames selected in a time interval of seven minutes. We chose to apply the median because of its efficacy in reducing the impulse noise \citep{gonzalez2002digital}.
%Figure~\ref{fig:noise} shows the value of the polarisation noise level ($\sigma$) for the single frame of  \FeI~6302~\AA\ line pair (top panel) and the median value of \CaII~8542~\AA\ (bottom panel).  The polarisation noise is shown for two different regions, labelled with \textit{in} and \textit{out}. These two regions are highlighted by yellow contours in Figure~\ref{fig:fe_flux}a.
We estimated the noise levels ($\sigma$) of Stokes Q, U, and V for the wavelength positions closest to the continuum as the standard deviation of the polarisation signals normalised to the continuum intensity, taking into account all the pixels of the SGS. For \FeI~6301-6302~\AA\, $\sigma$ is $1.5\times 10^{-3}$ in Q, U, and V. For \CaII~8542~\AA\ $\sigma$ is $1.2\times 10^{-3}$ in V and $1.0\times 10^{-3}$ in Q and U. These polarisation noise values are typical of ground-based filter polarimeters \citepads{2015SSRv..tmp..115L}. They can be compared with the signal values of Figure~\ref{fig:magneto}.} \secgrass{Example profiles of \CaII~8542~\AA\ are displayed in Figure~\ref{fig:ex_prof} for three different points within the SGS indicated as 1, 2, and 3 in Figure~\ref{fig:magneto}b.}

At the same time as CRISP, we observed in the blue part of the visible spectrum with CHROMIS, recording along the following spectral lines:
\begin{itemize}
\item[•]\CaIIK, at 21 equidistant positions between 3933.1 and 3934.3~\AA\ with $\Delta \lambda=$ 59~m\AA;
\item[•]\CaIIH, at 21 equidistant positions between 3967.9 and 3969.1~\AA\ with $\Delta \lambda=$ 59~m\AA;
\item[•] plus one continuum point at 3999.9~\AA.
\end{itemize} 
The entire \CaIIHK\ profile is scanned in 15~s. At 3933~\AA\ the pixel size is 0\farcs0375 and $R \approx 3.2 \times 10^{4}$. %
Due to the similar appearance of the \CaIIHK\ doublet lines, we have decided to show only \CaIIK\ which forms higher in the chromosphere \citepads{1975ApJ...199..724S}.  

The data were reduced using the CRISPRED pipeline \citepads{2015A&A...573A..40D} and CHROMISRED pipeline \citepads{2018arXiv180403030L} for CRISP and CHROMIS data respectively. 
Image restoration was performed by employing the Multi-Object Multi-Frame Blind Deconvolution technique
\citepads[MOMFBD,][]{1994A&AS..107..243L, 2005SoPh..228..191V}. 

Due to the higher spatial resolution of CHROMIS, we have scaled up the CRISP dataset to the CHROMIS pixel size and spatially aligned it to match the CHROMIS FOV by cross-correlating the wide-band images. However, all the calculations involving polarimetry have been performed on non-aligned data, to avoid degradation of the signal-to-noise ratio. 

During the same day, part of the SGS was also observed by the Solar Optical Telescope 
\citepads[SOT,][]{2008SoPh..249..167T} 
on board Hinode
\citepads{2007SoPh..243....3K}.
We employed level 2 data obtained by the SOT spectropolarimeter (SP) running in fast map mode with pixel size of $0\farcs32$. These data were taken between 19:21 and 20:08~UT, so they are not co-temporal with SST observations. 

In addition, we used observations with the Helioseismic and Magnetic Imager 
\citepads[HMI,][]{2012SoPh..275..207S} and the Atmospheric Imaging Assembly 
\citepads[AIA,][]{2012SoPh..275...17L}
on board of the Solar Dynamics Observatory 
\citepads[SDO,][]{2012SoPh..275....3P}
to study the time evolution of the SGS  during its transit from the limb to the disk centre. The full-disk image of the SDO instruments allows us to observe the entire SGS with a spatial resolution of 1$\arcsec$.

%******************************************************************************************
\section{Results}\label{sec:res}

\begin{figure*}
\includegraphics[scale=1]{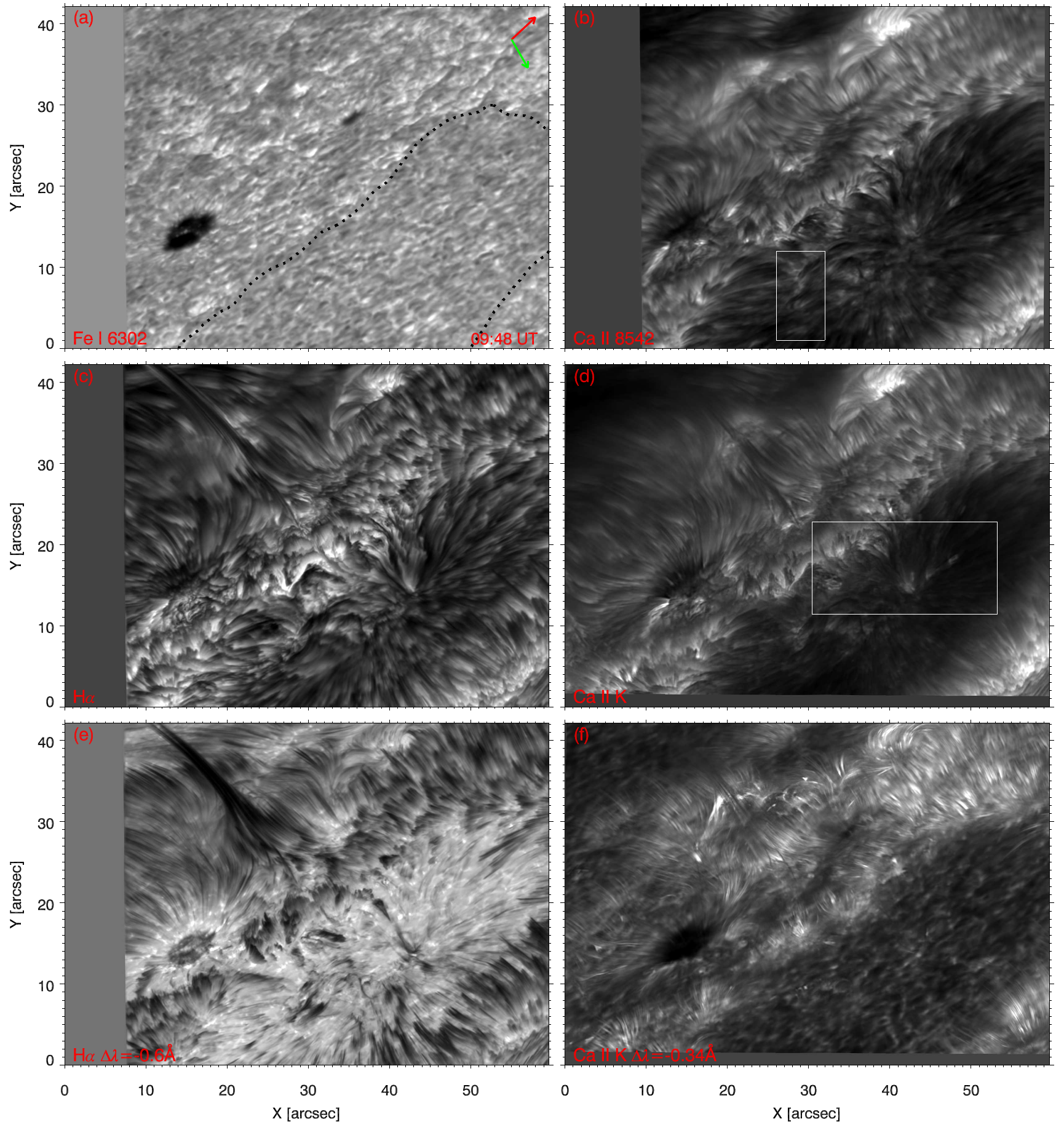}\caption{Nominal line core intensity map of \FeI~6302~\AA\ (a), \CaII~854.2~\AA\ (b), \halpha\ (c) and \CaIIK\ (d) and blue wings intensity of \halpha\ (e) and \CaIIK\ (f) at $\Delta \lambda=-0.6$~\AA\ and $\Delta \lambda=-0.34$~\AA\ respectively. The images are spatially and temporally co-aligned according to the CHROMIS field of view. \secgrass{The black dotted curve identifies the bright network.} The white rectangle in panel b corresponds to the one shown in Figure~\ref{fig:magneto}b while the rectangle in panel d highlights the region shown by Figure~\ref{fig:zoomin}. The red arrow indicates the north and the green arrow the disk centre. \secgrass{The temporal evolution is shown in a movie in the online edition.}} \label{fig:linecores}
\end{figure*}

Figure~\ref{fig:linecores} displays the nominal line cores of \FeI~6302~\AA\ (a), \CaII~8542~\AA\ (b), \halpha\ (c) and \CaIIK\ (d) and the blue wings of \halpha\ (e) and \CaIIK\ (f). All the images are rotated according to the CHROMIS FOV (panels d and f). The north direction is indicated by the red arrow and the disk centre by the green arrow.  Panel a shows  photospheric faculae forming a circular bright network, which appears elliptical because of the projection effects. \secgrass{The black dotted curve} helps the reader to locate the bright network.
This structure belongs to an active region featuring a small and a large pore. 
Panels b-d show chromospheric fibrils seemingly rooted in the bright annular network, and converging towards the centre of the SGS. The fibrils in the three chromospheric lines (b-d) show a very similar pattern and single fibrils can be distinguished in all the three panels. 
The fibril canopy forms a darker area against the active region background and highlights a structure of supergranular size ($\sim 50\arcsec$). Fibrils of the QS look less combed on the southern and eastern sides of the SGS with respect to those on the northern and western sides. \secgrass{They also exhibit a different apparent length.}
The entire structure reminds a rosette, except a smaller region in the S-E of the SGS, where fibrils are arranged as a chain. Here, one end of the fibrils is rooted in the annular network while the other lies inside the SGS. This second footpoint location is highlighted in panel b by a white rectangle. This chain is present during the entire observation but it is well visible only in  \CaII~8542~\AA\ dataset. %Unlike in the rest of the SGS, \halpha\ image looks here very different from the other chromospheric lines.

%\begin{figure}
%\includegraphics[scale=1]{fig/fig_wings.eps}\caption{Intensity map of  \halpha\ at $\Delta \lambda=-0.6$~\AA\ (a) and \CaIIK\ at $\Delta \lambda=-0.34$~\AA\ (b). The red arrow indicates the north and the green arrow the disk centre.}\label{fig:wings}
%\end{figure}
The different appearance of the fibrils can be further appreciated in the blue wing of \halpha\ (Figure~\ref{fig:linecores}e). On the west side of the SGS, the dark fibrils look thinner and longer and they are rooted in the bright network. On the contrary, the fibrils look shorter and more clustered on the east side. 
In the blue wing of \CaIIK\ (Figure~\ref{fig:linecores}f) we observe bright and thin fibrils, which are known as straws
\citepads{2007ASPC..368...27R}
or slender \CaIIH\ fibrils 
\mbox{\citepads{2017ApJS..229...11J}}.
In our observations, these fibrils are rooted exactly above the bright granules of the AR (Figure~\ref{fig:linecores}a), in agreement with previous publications \citepads{2007ASPC..368...27R,2009A&A...502..647P,2017ApJS..229...11J}. Moving towards the line centre, the straws progressively lose their sharpness and become partially covered by dark and opaque fibrils. We do not note any remarkable morphological difference between straws of different sides. Straws have been suggested to be the tracer of the magnetic field direction \citepads[e.g.][]{1989ApJ...337..964S,2017ApJS..229...11J}, thus we would expect a rather vertical field all along the network ring surrounding the SGS.

\subsection{Magnetic topology}\label{subs:topology}

\begin{figure*}
\centering
\includegraphics[scale=1]{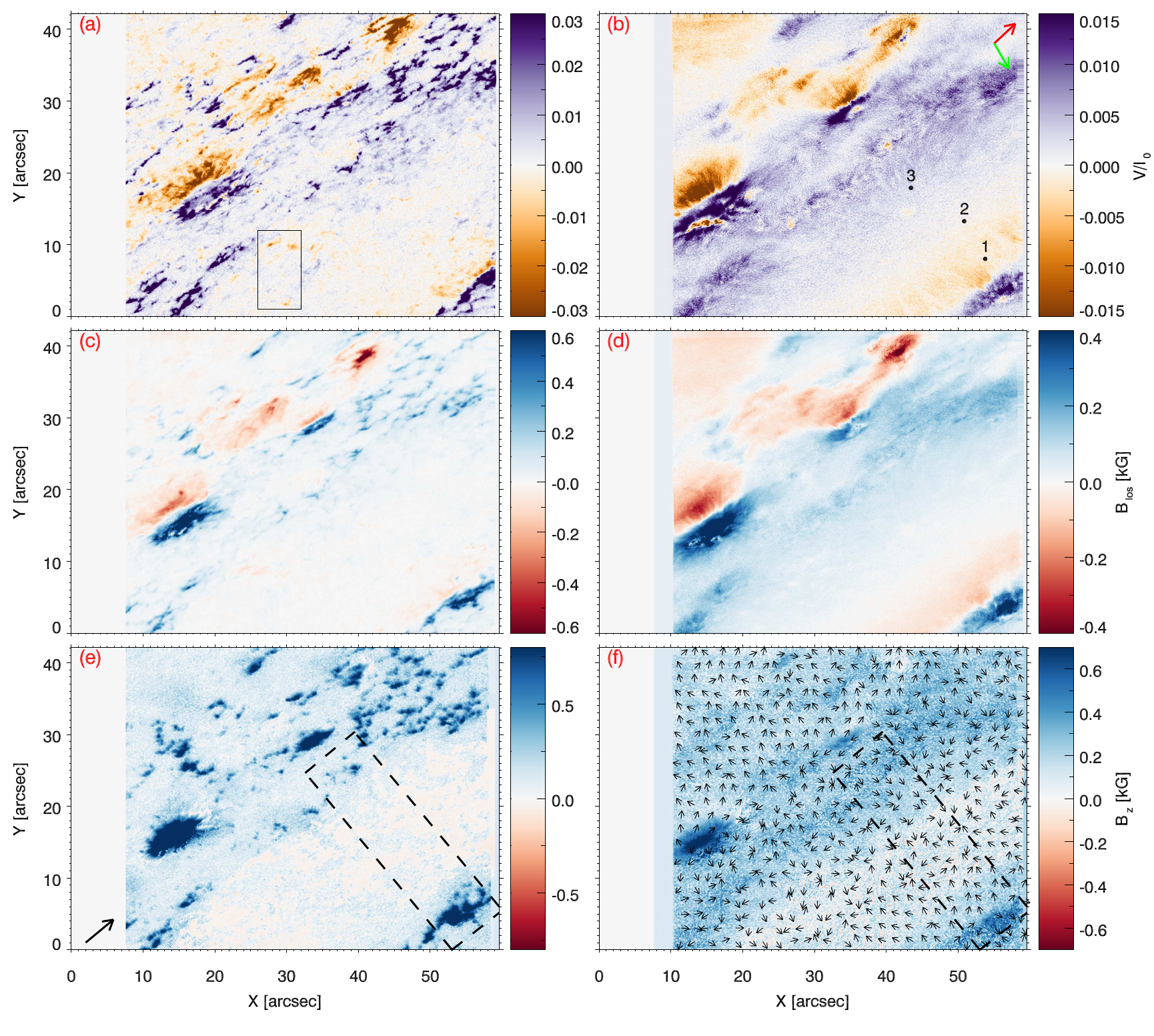}\caption{\grass{Polarisation and magnetic field maps: Circular polarisation of  (a) \FeI~6302~\AA\ 
at $\Delta \lambda=-0.08$~\AA\  and (b) \CaII~8542~\AA\ at $\Delta \lambda=-0.29$~\AA;
 Line-of-sight magnetic field strength obtained from (c) the \FeI~6301-6302~\AA\ and (d) \CaII~8542~\AA;
 Vertical magnetic field strength  for (e) the \FeI~6301-6302~\AA\ (f) and \CaII~8542~\AA. In panel b, the red arrow indicates the north and the green arrow the disk centre. \secgrass{The points indicated with 1, 2, and 3 refer to Figure~\ref{fig:ex_prof}.} The dashed rectangle and the black arrow of panel e refer to Figure~\ref{fig:incl}. The black arrows in panel f indicate the direction of the horizontal field.}}\label{fig:magneto}
\end{figure*}

\begin{figure}
\includegraphics[scale=1]{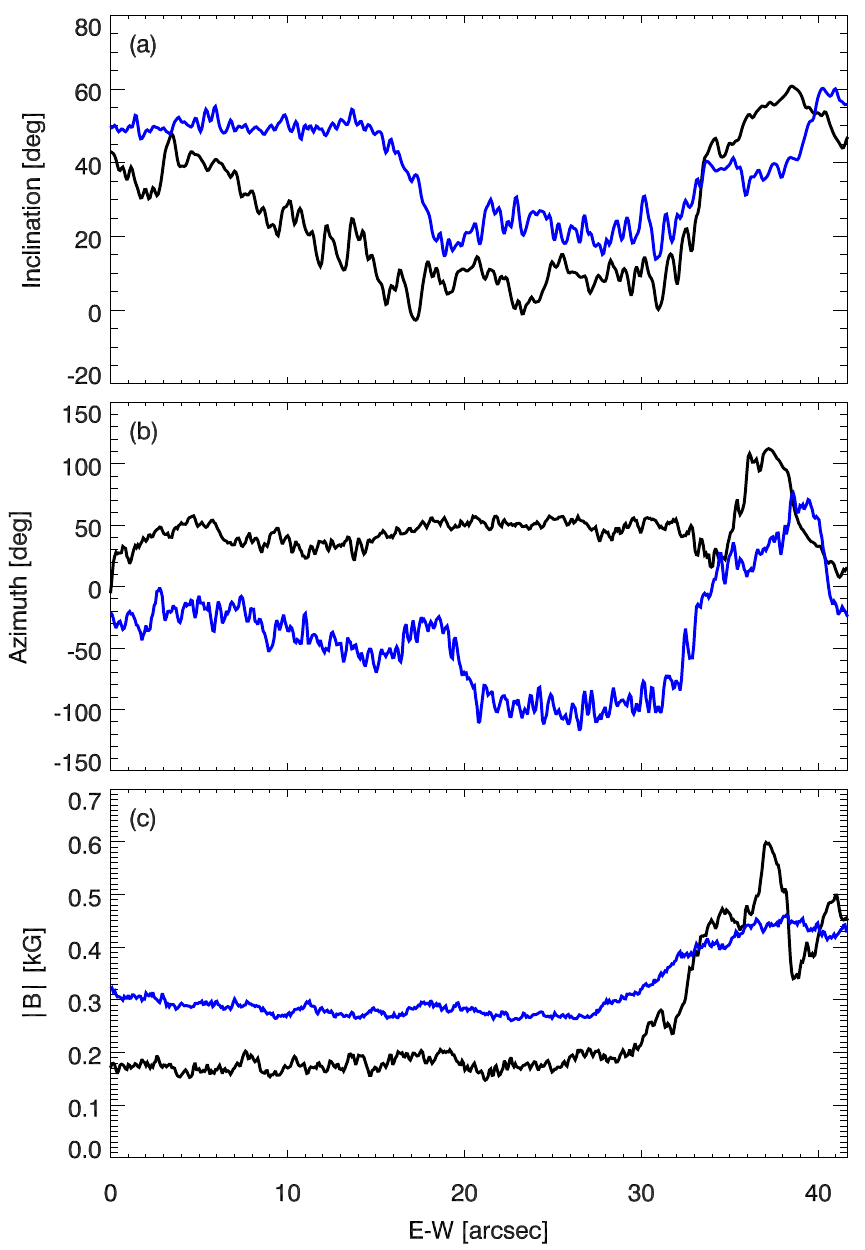}\caption{\secgrass{Value of the inclination (a), azimuth (b), and strength (c)} of the magnetic field obtained from \FeI~6301-6302~\AA\ (black curve) and \CaII~8542~\AA\ data (blue curve) averaged inside the dashed box of Figure~\ref{fig:linecores}e along the direction indicated by the black arrow of Figure~\ref{fig:linecores}e. Vertical magnetic field corresponds to an inclination of 90$^{\circ}$. The zero azimuth is defined with respect to the west direction (counterclockwise).}\label{fig:incl}
\end{figure}

\grass{To retrieve the photospheric magnetic field vector of the entire FOV, we performed simultaneous inversion of the \FeI~6301-6302~\AA\ lines using the MINE code (de la Cruz Rodriguez et al. in prep).
MINE is a parallel single component Milne-Eddington (ME) inversion with analytical response functions \citepads{1987ApJ...322..473S, 2007A&A...462.1137O}. It can include the 6301~\AA\ and 6302~\AA\ lines in one single inversion following the scheme presented by \citetads{2010A&A...518A...3O}. MINE adjusts the physical parameters
of the guessed input model in order to minimise the difference between the
observed and synthetic profiles using a standard Levenberg-Marquardt
algorithm. The synthetic profiles and the response functions are
convolved with the CRISP instrumental profile after each spectral
synthesis.
Here, we performed the inversion in two runs. In the first run, we assigned the same input model atmosphere to all the pixels. The atmospheric parameters are the total magnetic field strength, the magnetic field inclination, the azimuth, the line-of-sight velocity, the Doppler width, the line opacity, the damping parameter, and source function parameters. Macrotubulence is set to zero. We used a smoothed version of the output model to feed the second run, which produces the final result.}
%\begin{table}\caption{\grass{Atmospheric parameters of the input model atmosphere used for the Milne-Eddington inversion. Total magnetic field strength B, magnetic field inclination $\theta$ and azimuth $\psi$, line-of-sight velocity V$_{\textrm{los}}$, Doppler width $\Delta \lambda_{\textrm{D}}$, line opacity $\alpha$, damping parameter a, and source function parameters S$_0$ and S$_1$.}}\label{tabl:me}
%\centering
%\bgroup
%\def\arraystretch{1.5}
%\begin{tabular}{cccc}
%\hline
%\hline
%\textbf{B [G]} & \textbf{$\mathbold{\theta}$~/~$\mathbold{\psi}$ [rad]}  & \textbf{V$_{\mathbold{\textrm{los}}}$ [\kms]}  & $\mathbold{\Delta \lambda_{\textrm{D}}}$ \textbf{[\AA]} \\
%300 &  0.78 / 0.39  &       1 &     0.04 \\   
%
%$\mathbold{\alpha}$ & \textbf{a} \textbf{[$\mathbold{\Delta \lambda_{\textrm{D}}}$]} & \textbf{S}$\mathbold{_0}$ & \textbf{S}$\mathbold{_1}$ \\
% 40 &      0.3 &      0.4 &       1.9  \\
%\hline
%\hline
%\end{tabular}
%\egroup
%\end{table}

\grass{The chromospheric magnetic field is obtained using the weak field approximation from the temporal median of nine \CaII~8542~\AA\ frames. We applied the relations derived in \citetads{degl2006polarization},}

\begin{align}
V(\lambda) &= - \frac{\lambda_0^2\,e_0}{4 \pi m_e c^2} \,\dfrac{\bar{g}}{\lambda-\lambda_0}\,B\, cos \theta \dfrac{\partial I}{\partial \lambda},\label{eq:vwfa}\\
U(\lambda) &=\frac{3}{4} \frac{\lambda_0^4\,e_0^2}{(4 \pi m c^2)^2} \,\dfrac{\bar{G}}{\lambda-\lambda_0}\, B^2 sin^2 \theta \, \dfrac{\partial I}{\partial \lambda} \, \sin 2 \psi, \label{eq:uwfa} \\
Q(\lambda)&=\frac{3}{4} \frac{\lambda_0^4\,e_0^2}{(4 \pi m c^2)^2} \,\dfrac{\bar{G}}{\lambda-\lambda_0}  \,B^2 sin^2 \theta \, \dfrac{\partial I}{\partial \lambda} \, \cos 2 \psi \label{eq:qwfa}
\end{align}
\grass{where $\lambda_0$ is the nominal spectral line centre, $e_0$ the electron charge, $m_e$ the electron mass, $c$ the speed of light, $\bar{g}$ the effective Land\'e factor, $B$ the magnetic field strength, $I$ the intensity, $\psi$ the azimuthal angle of the magnetic field, and $\theta$ the inclination. $\bar{G}$ is a function of the angular momenta of the transition levels. Its full expression can be found in \citetads{degl2006polarization}.
Equation~\ref{eq:vwfa} holds true along the entire spectral line profile, while Equation~\ref{eq:uwfa} and \ref{eq:qwfa} better applies in the wings. Therefore, we omitted the nominal line core in the calculations of the transversal component of the magnetic field. 
The two components of the magnetic field (longitudinal and transversal) and the azimuth are obtained by least-squares minimisation.}

The transversal component of the magnetic field is affected by the Zeeman ambiguity \grass{regardless of the technique employed to calculate the field. The ambiguity} has to be removed in order to rotate the reference system from line-of-sight (LOS) to local solar coordinates (LSC). 
\grass{In regions of relatively strong photospheric field, such as our AR and annular network, the ambiguity can be solved by using standard methods.} We applied the minimum energy method 
\citepads[MEM,][]{1994SoPh..155..235M}, using the implementation of \citetads{2014ascl.soft04007L}.
Solving the ambiguity in the portion of QS contained in the SGS is more challenging because the polarisation signal is lower than in the AR and there is no clear indication for a preferred field direction. Previous studies 
have shown that the field in the QS tends to be horizontal 
\citepads[eg.][]{2008ApJ...672.1237L, 2009A&A...506.1415B, 2011A&A...530A..51B,2012ApJ...751....2O}
, in particular
\citetads{2011A&A...530A..51B}
noticed that the stronger the magnetic field is, the larger is \grass{the angle between the solar surface and the field lines.}
Considering this remark, we have applied the following strategy: first, we have calculated the magnetic field vector in the LSC for the two possible azimuthal directions and, using these values, the two possible inclinations. Then, we have imposed these conditions to select the most likely value:
\begin{enumerate}
\item Observing that, for our large heliocentric angle, the horizontal magnetic field in the LOS component ($\textrm{B}^{\textrm{los}}_{\textrm{x}}$) is the closest in magnitude to the LSC longitudinal field ($\textrm{B}_{\textrm{z}}$), then, when $\textrm{B}^{\textrm{los}}_{\textrm{x}} > 500$~G we choose the azimuth that maximises the  inclination. 
\item \grass{Close to the limb, the largest part of circular polarisation signal is produced by horizontal magnetic field. Thus, the line-of-sight magnetic field (Figure~\ref{fig:magneto}c) provides an indication of the actual horizontal magnetic field direction in the LSC, without ambiguity. Since the magnetic field in the pore is positive and radial, we can infer that, if B is negative along the LOS, the horizontal field is directed from the disk centre towards the SGS centre. Otherwise, the field is directed from the limb towards the disk centre.}
Thus, for those pixels where $\textrm{B}^{\textrm{los}}_{\textrm{x}} < 500$, we impose that, if $\textrm{B}^{\textrm{los}}_{\textrm{z}}$ is positive, the azimuth, which in our case ranges between 0$^{\circ}$ and 180$^{\circ}$, is smaller than 90$^{\circ}$. 
\end{enumerate}
The results have been compared to the $\textrm{B}^{\textrm{los}}_{\textrm{z}}$ map of the same SGS observed by HMI five days later, which was close to disk centre. The general polarity distributions are consistent with each other for both the AR and the QS.

\grass{Figure~\ref{fig:magneto}a shows the Stokes \textit{V} signal of \FeI~6302~\AA. The signal is stronger along the boundaries of the SGS while there is a weaker pattern of internetwork within the SGS. 
The black inset indicates the second footpoint of the fibrils arranged as a chain. The Stokes  \textit{V} signal in this region is slightly higher than in the rest of the inner SGS, probably affecting the fibril arrangement.  Figure~\ref{fig:magneto}c shows the LOS magnetic field retrieved by ME inversion. The vertical component of magnetic field in the LSC is displayed in Figure~\ref{fig:magneto}e. From this map, we can read that the AR and the network ring around the SGS are positive and the SGS has a weak mixed polarity. 
We selected the pixels in the dashed box of Figure~\ref{fig:magneto}e and calculated the mean
 inclination by averaging the pixels along the direction shown by the black arrow. The black curve of Figure~\ref{fig:incl}a represents the average inclination of the SGS structure, with 0$^{\circ}$ equal to a horizontal magnetic field. The network ring is on average more vertical than the QS inside the SGS.
In the same way, we plotted the averaged azimuth in panel b (black curve). The zero coincides with the longer axis of the dashed box directed westwards.} \secgrass{Panel c shows the average value of the magnetic field strength calculated as in panel a and b. The photospheric field (black) is constant within the SGS and rapidly increases at the supergranular boundaries.}

\grass{The same quantities of the left column of Figure~\ref{fig:magneto} are shown for \CaII~8542~\AA\ on the right. Panel d displays the longitudinal magnetic field component obtained in weak field approximation. 
To choose the direction of transversal field, we applied the second argument employed for the \FeI~6302~\AA\ azimuth solution.
The correct azimuth is used to rotate the magnetic field vector in the local solar reference frame. Figure~\ref{fig:magneto}f shows the vertical magnetic field in the LSC (B$_{\textrm{z}}$).}

\grass{We calculated the chromospheric field inclination in the same way as the photosphere and displayed it as a blue curve in Figure~\ref{fig:incl}a.  The right side of the curve (west side of the network) corresponds to an extended concentration of strong magnetic field. Here, the photospheric magnetic field is more vertical. The blue curve in Figure~\ref{fig:incl}a shows that the magnetic field obtained by the weak field approximation tends to be rather horizontal on the centre of the SGS, while it is more inclined on the boundaries. 
In panel b, we show  the average azimuth (blue curve). On the west side of the SGS, this is -110$^{\circ}$ while on the east side it is about -20$^{\circ}$. The azimuth is also indicated by the black arrows in Figure~\ref{fig:magneto}f. These arrows show that on the west side of the box, a part of the field lines converges from the north towards the centre of the SGS with an azimuth close to -90$^{\circ}$. Another set of field lines converges from the west with an azimuth of roughly -135$^{\circ}$. This is in agreement with the direction of the corresponding fibrils (compare to Fig.~\ref{fig:linecores}). 
Thus, the average azimuth ($\sim$ -110$^{\circ}$) in the west side of the box (Fig.~\ref{fig:incl}b) represents the average between these two main directions. The azimuth values on the east side of the box ($\sim$ -20$^{\circ}$) and  on the west side ($\sim$ -110$^{\circ}$) indicate that the horizontal field tends to point towards the inner part of the SGS.} \secgrass{The blue curve of panel c represents the average chromospheric magnetic field strength. Its value above the SGS is rather constant and, similarly to the photosphere, it increases above the boundaries.}

\begin{figure}
\includegraphics[scale=1]{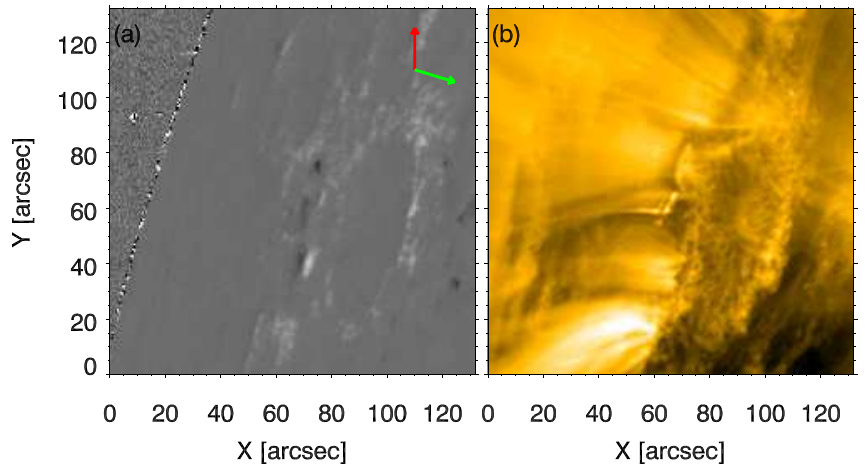}\caption{\textit{Left:} SDO/HMI line of sight magnetogram with a larger field of view respect to Figure~\ref{fig:magneto} and different orientation. The red arrow indicates the north and the green arrow the disk centre. \textit{Right:} SDO/AIA 171~\AA\ image with the same FOV of panel a. Both images are co-temporal to images in Figure~\ref{fig:linecores}.}\label{fig:sdo}
\end{figure}

A co-temporal observation with SDO (Figure~\ref{fig:sdo}) shows that no structure is visible above the SGS in 171\AA, except for a weakly bright structure following the fibrils arrangement, which might be a hint of a low lying canopy. The network is bright and several vertical plumes fan out from the outer part of the ring, where the field is more concentrated. 
%%
%\citetads{2016ApJ...818..203W}
%%
%noticed that coronal plumes tend to form above unipolar flux concentration created by the convergence motion of supergranules. They are also often associated to minority polarity inside the network that are not visible in our magnetogram (Figure~\ref{fig:magneto}-a).

\subsection{Flux imbalance}

%dalle immagini in corona sembra suggerire che gran parte dle flusso vada su
%possiamo dunque calcolare il flusso...
%considerazioni
%usando le stesse maschere possiamo calcolare l asimmetria del granulo non pi' solo in qs.
%posso metterlo in relazione con le fibrils

The \grass{vertical magnetic} field allows us to compute the total \grass{photospheric} flux inside the SGS and along its annular network.
The general magnetic flux expression,
 \begin{equation}
  \Phi(\textrm{B}) = \int_{S} \textrm{\textbf{B}}\cdot \textrm{d}\textrm{\textbf{s}}
 \end{equation}
 can be discretised into an expression for the positive flux, $+\Phi(\textrm{B})$, and another for the negative flux, $-\Phi(\textrm{B})$,
\begin{equation}
\pm \Phi(\textrm{B})= \sum_{x}\sum_{y} \pm \textrm{B}_\textrm{z}(\textrm{x,y}) \,\, \Delta  \textrm{A}
\end{equation}
where $\Delta  \textrm{A}$ is the pixel area. The positive (negative) flux is calculated by summing over all the pixels where $\textrm{B}_\textrm{z}$ is positive (negative).

\begin{figure}
\includegraphics[scale=1]{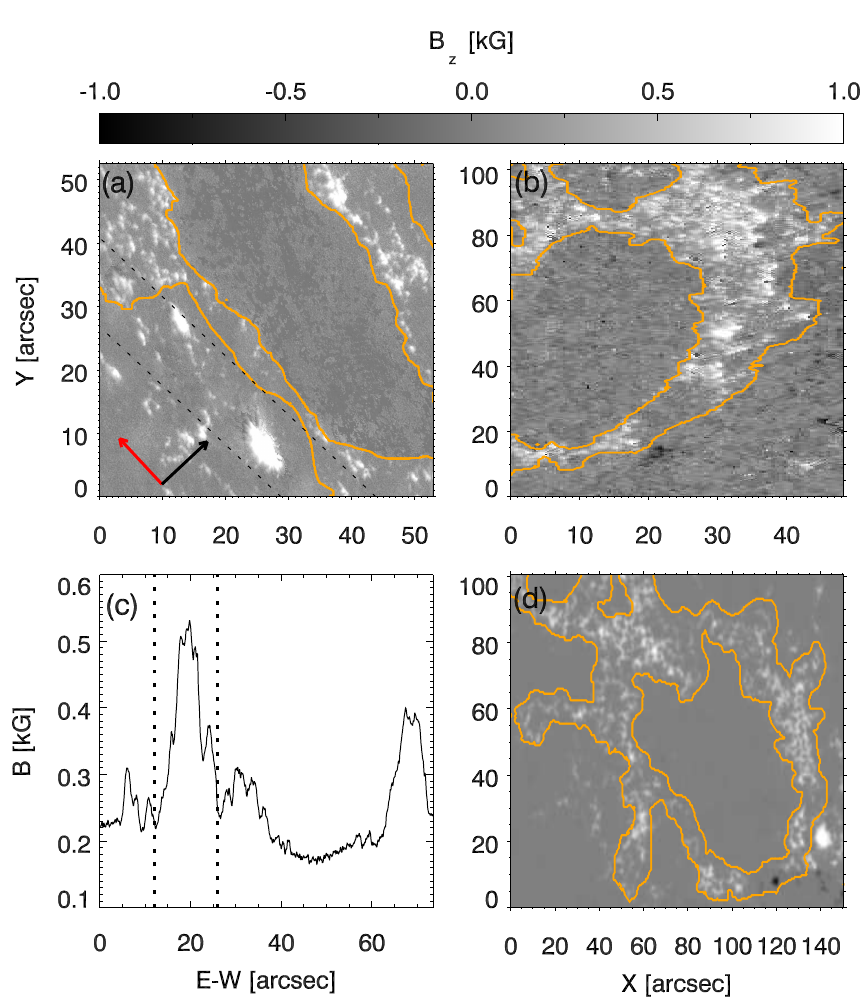}\caption{Vertical magnetic field map obtained with CRISP (a), HINODE/SP (b) and SDO/HMI (d). The red arrow indicates the north and the black arrow the E-W direction. In panel b and d, the north coincides with the top side of the image. The orange contour indicates the masks employed for the flux calculation. Panel c shows the average magnetic field strength along the E-W direction calculated from the CRISP data. The dotted lines in panels a and c higlight the region containing the pores.}\label{fig:fe_flux}
\end{figure}

The vertical magnetic field in Figure~\ref{fig:magneto}a has been aligned to the FOV of CHROMIS to match the intensity maps of Figure~\ref{fig:linecores}, but to calculate the flux we used the non-aligned longitudinal field map, shown in Figure~\ref{fig:fe_flux}a. However, the FOV of CRISP does not capture the entire SGS and the northern and southern sides of the cell are missed.

Therefore, we have decided to employ magnetograms of the same SGS observed by other instruments.
Hinode/SP data \grass{display a large fraction of the SGS ten hours after our observing run at the SST and show the missing structure of the CRISP data.}
To avoid projection effects, we also used the HMI magnetogram which was obtained close to disk centre on April 25. After five days, we still observe a similar unipolar region, but it is impossible to directly compare with SST data.
For all the data, we have manually created a mask to choose the inner SGS and another one to select the annular network at the boundaries. Both masks are indicated by orange contours in Figure~\ref{fig:fe_flux}a, b and d. As can be seen in panel a, the two pores within the dotted lines were intentionally excluded from the outer mask.
Table~\ref{tab:flux} summarises the results of the flux calculations. The \textit{in} superscript stands for the QS region within the SGS while \textit{out} is for the area enclosed by the two orange contours. 

Although we cannot compare the results inferred for each dataset because of the different FOV and resolution, we have nonetheless three independent measurements  of flux disequilibrium. The annular network has a strong positive flux that is not balanced by the negative flux of the entire SGS. In contrast, the positive and negative fluxes are closer to equilibrium in the inner region, although we observe an excess of positive polarity. 

The different order of magnitude obtained from the inner flux calculations, is likely due to the different spatial resolution of each instrument. The HINODE/SP network flux has the same order of magnitude as the lower-spatial-resolution SDO/HMI, but we have to consider that HINODE/SP FOV lacks a non-negligible fraction of the eastern side of the SGS. Therefore we expect the actual total value to be higher, as well as for the SST/CRISP network flux. 

\grass{Considering line core intensity filtergrams in \CaII~8542~\AA, we can choose a smaller mask enclosing the region where fibrils are bending towards the SGS centre.} This choice (not shown), however, decreases the value of the positive flux of the 20\% only, which is insufficient to balance the positive annular ring flux.

All the three cases (panels a, b, and d) show that the negative and positive field of the entire SGS are significantly unbalanced and the small negative polarity patches in the inner region are not enough to balance the positive network field.
\grass{The SDO/HMI coronal image in 171 Å shown in Fig.~\ref{fig:sdo} suggests that part of the positive flux might follow the locally open field lines rooted in the network and connect with negative polarity outside the FOV.}

There is an asymmetry in the magnetic field concentration between the east and west side of the SGS. This asymmetry is exemplified in Figure~\ref{fig:fe_flux}, where we plot the longitudinal magnetic field averaged along the N-S direction (red arrow in Figure~\ref{fig:fe_flux}a) as a function of E-W direction (black arrow). The increasing values on the x-axis goes from E to W. The region between the two dotted lines indicates the location of the pores. The network field enhancement is visible around X=20$\arcsec$ on the east side and around X=67$\arcsec$ on the west side. The east side is on average weaker and less concentrated. The same asymmetry was reported by \citetads{2015A&A...579L...7L} in a statistical study of more than 3000 supergranules observed close to disk centre. The different appearance of the SGS fibrils shows the same E-W asymmetry.
According to the scenario of the magneto-acoustic origin of type-I spicules \citepads[e.g.][]{2004Natur.430..536D,2007PASJ...59S.655D}, the length of the SGS fibrils should be dependent on the magnetic field inclination \citepads{1984A&A...132...45Z, 2006ApJ...648L.151J}.
Consequently, we would expect a more horizontal field on the west side, where the fibrils look longer.
\grass{The blue curve of Figure~\ref{fig:incl}a does not contradict this scenario. However, close to limb it is impossible to discern whether there is an actual physical correlation between the two asymmetries or whether it is a projection effect.}
\mbox{}\\
\begin{table*}\caption{Positive and negative flux calculated  inside the SGS ($\Phi^{in}_B$) and along the annular network ($\Phi^{out}_B$) for each of the instruments shown in Figure~\ref{fig:fe_flux}a, b, d. A$^{\textrm{in}}$ represents the total amount of positive (+) and negative (-) pixels in the inner mask, multiplied by the single pixel area. The same for A$^{\textrm{out}}$ in the outer mask.}\label{tab:flux}
\centering
\bgroup
\def\arraystretch{1.5}
\begin{supertabular}{l c c c c c c c c}
\hline
\hline

\multirow{2}{*}{Instrument} & \multicolumn{2}{c}{$\Phi^{\textrm{in}}_B$ [Mx]} & \multicolumn{2}{c}{$\Phi^{\textrm{out}}_B$ [Mx]} & \multicolumn{2}{c}{A$^{\textrm{in}}$ [cm$^{2}$]} & \multicolumn{2}{c}{A$^{\textrm{out}}$ [cm$^{2}$]}\\
 & + & - & + & -& + & -& + & -\\
\hline
SDO/HMI   & $4.6\cdot10^{23}$ & $1.1\cdot10^{23}$ & $5.4\cdot10^{25}$ & $1.1\cdot10^{22}$ & $8.9\cdot10^{18}$ & $5.2\cdot10^{18}$ &$2.3\cdot10^{19}$& $5.2\cdot10^{17}$\\

Hinode/SP & $2.7\cdot10^{24}$ & $1.8\cdot10^{24}$& $4.2\cdot10^{25}$& $1.5\cdot10^{23}$ & $3.9\cdot10^{18}$ & $3.4\cdot10^{18}$ & $7.7\cdot10^{18}$ & $8.8\cdot10^{17}$\\

SST/CRISP (\FeI)& $5.8\cdot10^{25}$ & $1.3\cdot10^{25}$ & $2.5\cdot10^{26}$ & $7.57\cdot10^{21}$ & $3.2\cdot10^{18}$ & $2.5\cdot10^{18}$ & $4.3\cdot10^{18}$ & $7.9\cdot10^{16}$\\

%SST/CRISP (\CaII) & $2.2\cdot10^{26}$ & $9.6\cdot10^{24}$ & $2.2\cdot10^{26}$ & $1.23\cdot10^{23}$ & $4.5\cdot10^{18}$ & $1.1\cdot10^{18}$ & $4.2\cdot10^{18}$ & $2.1\cdot10^{17}$\\

\hline
\hline
\end{supertabular}
\egroup
\end{table*}
\mbox{}\\

\subsection{LOS Velocity}\label{sec:los}
\grass{The online animation of Fig.~\ref{fig:linecores} shows a dynamic atmosphere above the SGS. Thanks to multi-wavelength observations, we can provide velocity maps at different atmospheric heights. }

\begin{figure}
\includegraphics[scale=1]{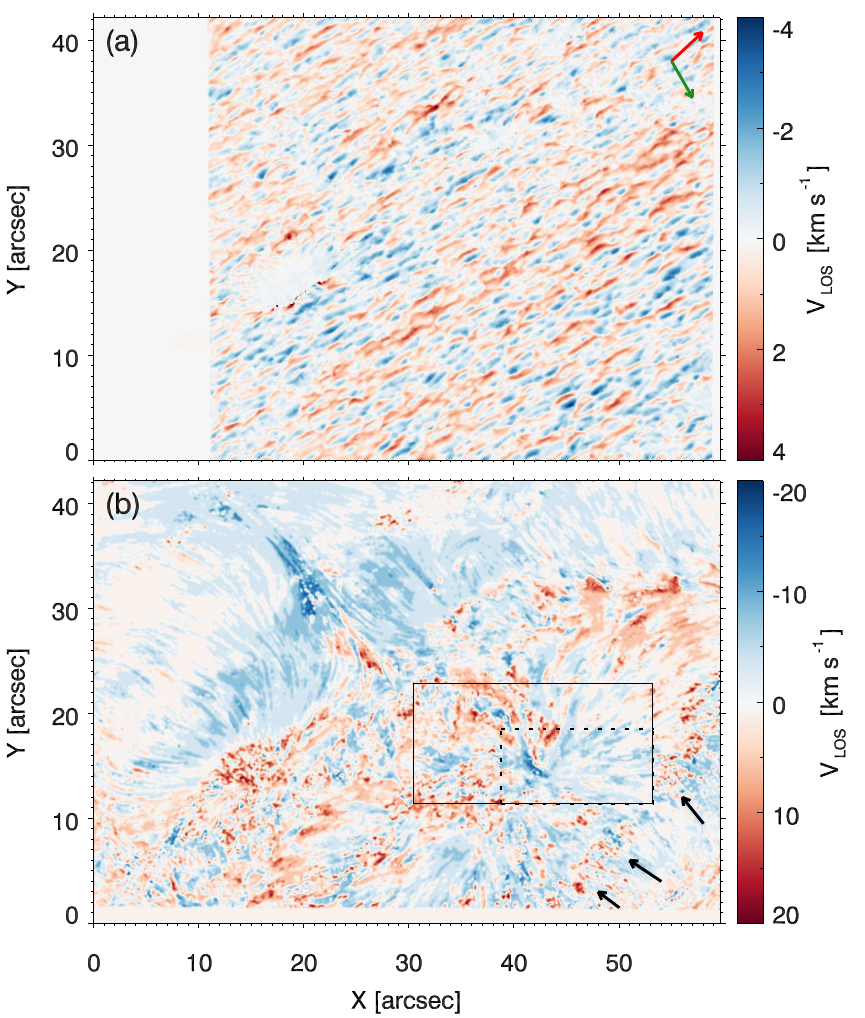}\caption{\textit{Upper:} photospheric LOS velocity obtained with Milne-Eddington inversion of \FeI~6301-6302~\AA. \textit{Lower:} chromospheric LOS velocity obtained from the K$_3$ features Doppler shift. The black solid rectangle highlights the region shown in Figure~\ref{fig:zoomin}, the dotted rectangle the region in Figure~\ref{fig:evol_vel} and~\ref{fig:clustering}.}\label{fig:vel}
\end{figure}

%
%los vel photo: come si ottiene
Figure~\ref{fig:vel} displays the LOS velocity map of the photosphere (panel a) and chromosphere (panel b) at 09:48~UT. The photospheric map is obtained from ME inversion and it was calibrated using as reference the velocity inside the pore, similarly as proposed by \citetads{1977ApJ...213..900B}. % and compensates for the blue shift produced by solar rotation. 
%
%cosa si vede.
The LOS velocity map mostly \grass{shows the LOS projection of the granular flows}. We do not observe any anomalies neither for the time step shown in Figure~\ref{fig:vel}a, nor for the entire time series. 
%
%los vel chromos: come si ottiene
Panel b displays the LOS velocity obtained from the Doppler shift of the \CaIIK\ core (K$_3$), which is a good diagnostic tool for the upper chromospheric velocity \citepads{2018A&A...611A..62B}.
%
%xke abbiamo scelto di utilizzare k3
We preferred to compute the LOS velocities from the K$_3$ peak because \CaIIK\ is less affected by thermal broadening than \halpha\ and it does not suffer of the inverse C-shape that concerns \CaII~8542~\AA. 
%
%come si costruisce l algoritmo e come definisco i tre k.
To identify the K$_3$ feature, we developed an algorithm that automatically searches for the line core of \CaIIK. It uses the definition of K$_3$ based on a typical QS profile of \CaIIK\, where the K$_3$ feature corresponds to the intensity minimum confined between two emission peaks, known as K$_{2{\textrm{v}}}$ on the blue wing and K$_{2{\textrm{r}}}$ on the red wing \citepads{1904ApJ....19...41H}. 
However, this definition is not suitable for some spectral profiles, as those emerging from strong magnetic field regions like the pores and the annular network which exhibit a  single emission peak. There, the algorithm identifies K$_3$ as the wavelength corresponding to the minimum intensity.
We calculated the chromospheric LOS velocity using the Doppler shift of K$_3$, wavelength-calibrated with the NSO/Kitt Peak FTS atlas \citepads{fts}.
\secgrass{To perform a precise velocity calibration, we also considered the centre-to-limb variation of the LOS velocity. 
From the nominal line centre value of the atlas, we estimated a velocity offset of about -600~\ms, which is comparable to the results  of \citetads{2011A&A...528A.113D}.}
All the velocities are also corrected for the blue-shift produced by the solar rotation ($\sim1.8$~\kms).

%
%cosa si vede in generale:difficiel interpretaz
Interpretation of the entire map is difficult due to the large observing angle and the different inclination of each fibril. It is not possible to detect a  clear bulk motion along them, but as a general trend, we notice that the structure closer to disk centre is more blue-shifted while, on the other side, profiles tend to be more red-shifted. %This is in agreement with the outwards horizontal motions usually observed in the photosphere of supergranules and it could also be compatible with a siphon flow scenario \citepads{1968MitAG..25..194M}.
The black arrows in Figure~\ref{fig:vel}b mark the transition between the apparently vertical and bright fibrils and the darker and horizontal ones, as we observed in \CaII~8542~\AA\ and \CaIIK\ line cores.

 \begin{figure*}
 \sidecaption
  \includegraphics[scale=1]{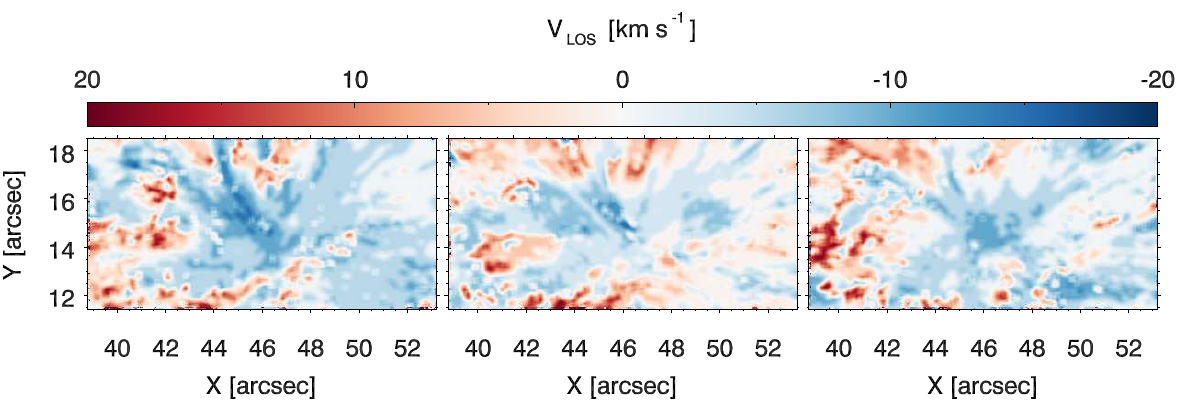}
     \caption{Enlargement of the dotted rectangle of Figure~\ref{fig:vel}b at three different time steps: 09:43~UT (\textit{left}), 09:52~UT (\textit{middle}),  and 09:59~UT (\textit{right}).}
     \label{fig:evol_vel}
\end{figure*}

%
%cosa si vede al centro
In the convergence point of most of the fibrils (around X=42$\arcsec$, Y=14$\arcsec$), a significantly blue-shifted structure appears. Comparing with Figure~\ref{fig:linecores}b, c, and d, we recognise part of a fibril. 
%
%dire che si vede sempre durante tti e 20 minuti
This strong blue-shift is visible during the entire observation ($\sim$20 min) at the same location (see Figure~\ref{fig:evol_vel}) and its shape changes during the observation.
%
%possibile upflow e downflow
It is non-trivial to establish whether this blue-shift is an actual upflow or a downflow observed along a structure that is inclined towards the limb.
%
%\begin{figure}
%\includegraphics[scale=1]{fig/fig_evol_vel.eps}\caption{Enlargement of the dotted rectangle of Figure~\ref{fig:vel}-b at three different time steps: 09:43~UT (a), 09:52~UT (b),  and 09:59~UT (c)} \label{fig:evol_vel}
%\end{figure}
%

\begin{figure}
\includegraphics[scale=1]{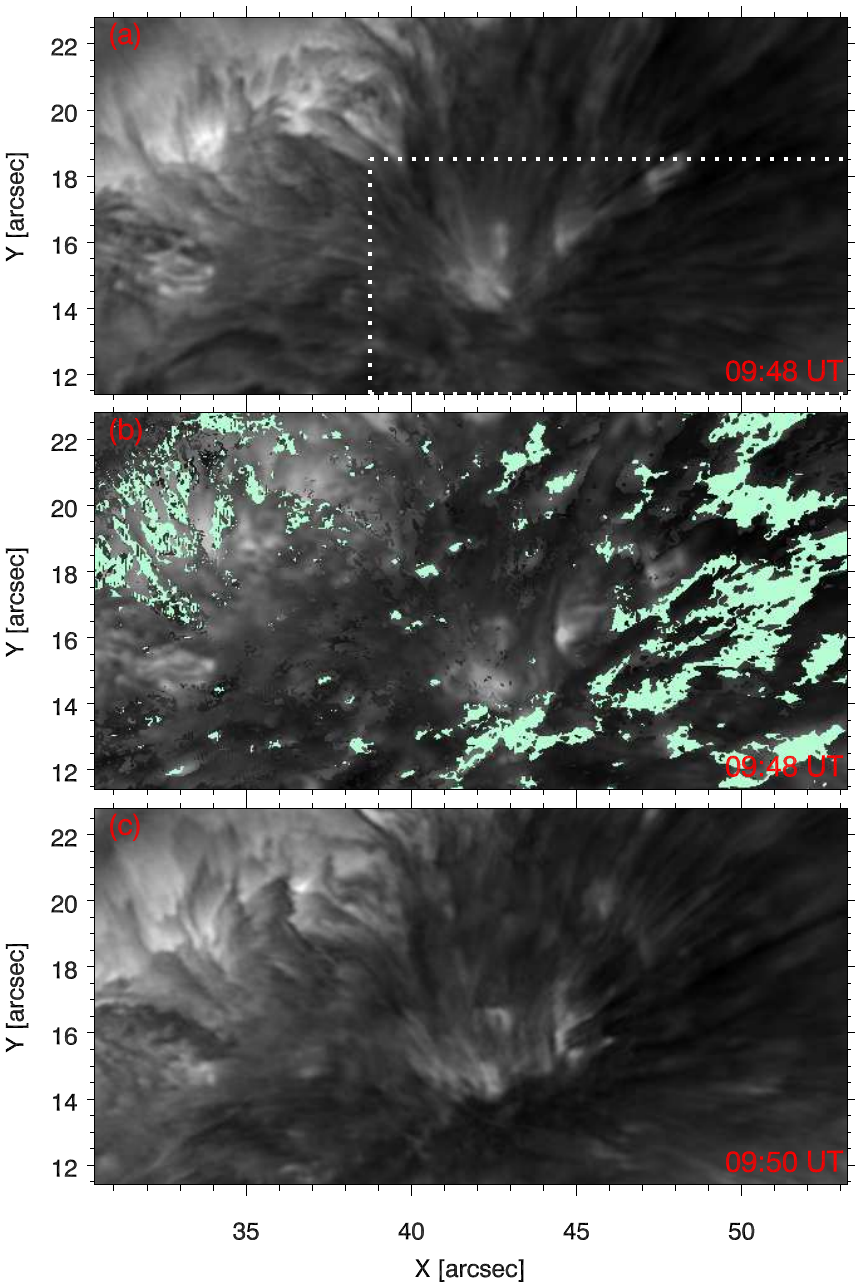}\caption{Close-up of the white rectangle shown in Figure~\ref{fig:linecores} of the \CaIIK\ nominal line centre (a) and the K$_{2\textrm{r}}$ intensity (b). The light green mask represents the pixels where the K$_{2 \textrm{r}}$ feature cannot be detected. Panel c shows the same field of view in the nominal line centre of \CaIIK\ at a different time (09:50~UT). All the images are corrected with a gamma factor of 0.5.}\label{fig:zoomin}
\end{figure}

%
%vedere come questa parte di blueshift appare in intensita: quindi birrlante rispetto al surrounding e abbiamo una stuttura apaprentemente verticale
Figure~\ref{fig:zoomin} shows a close-up of the centre of the SGS, which is enclosed by a white rectangle in Figure~\ref{fig:linecores}d.
Panel a shows the intensity in the nominal line centre of \CaIIK\ at the same time step used to calculate the Dopplergram shown in Figure~\ref{fig:vel}. A bright structure appears at the centre. 
Panel b displays the intensity map corresponding to the wavelength position that the algorithm identifies as K$_{2\textrm{r}}$ peak, at the same time step of panel a. 
\grass{Most of the pixels have a well-defined \CaIIK\ intensity profile and the K$_{2\textrm{r}}$ peak is successfully identified. However, there are some pixels where they have eluded detection (masked in light green).
The K$_{2\textrm{r}}$ intensity shows the same enhancement of panel a.
To exclude} that the chosen frame has a special appearance, we display the same region at a different time step (panel c). Here we observe an apparent vertical fibril arrangement with a brighter bottom. The online animation of Figure~\ref{fig:linecores} shows that this intensity enhancement in the core of the SGS is always present. 

\begin{figure}
\includegraphics[scale=1]{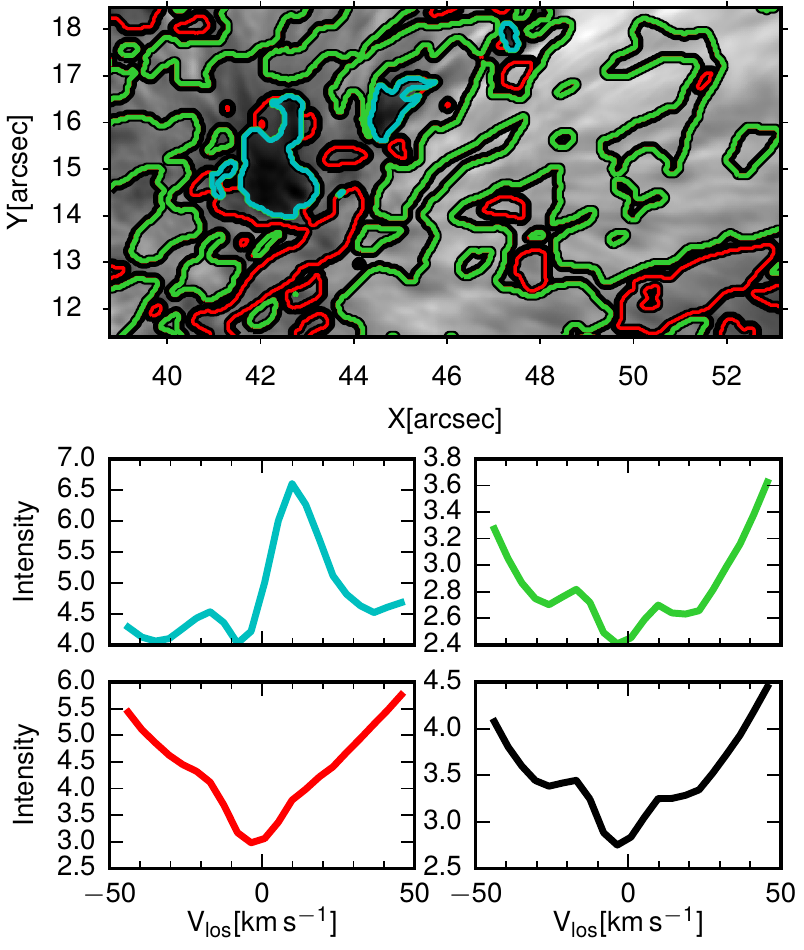}\caption{K-mean clustering of the spectral profiles of the inset of Figure~\ref{fig:zoomin}a. Each contour delimits the area where the profiles are closer to the corresponding coloured average profile. The upper image has reversed intensity and the intensity scales of the last four panels is in units of $10^{-7}$~$\mathrm{erg} \, \mathrm{cm}^{-2} \, \mathrm{s}^{-1} \, \mathrm{Hz}^{-1} \, \mathrm{ster}^{-1}$. The Y-axes have different scales.}\label{fig:clustering}
\end{figure}

%
%e' legittimo far vedere che tipo di profili caratterizzano la parte brillante delle fibirls.
To understand the origin of the bright SGS centre, we analysed the profiles \grass{emerging from the region enclosed by the dotted rectangle in Fig.~\ref{fig:zoomin}a}
We applied the k-mean algorithm to cluster the spectral profiles in four different typical profiles. 
%
%e quindi plottiamo clustering di una sub regione. questo xke l intero zoom di figura 8 comprende anche una parte di network dove il campo e' forte a cui nn siamo interessati. 
%We did not select the entire FOV of Figure~\ref{fig:zoomin} because it contains part of the annular network in which we are not interested. We chose instead the region enclosed by the dotted rectangle in Figure~\ref{fig:zoomin}a. 
%
%vediamo che il profilo dell'intera zona e' caratterixzato da una forte asimmetria tra k2 e k3. all interno di questa zona abbiamo sia una zona di forte blushift sia di redshift entrmabe che caratterizzate da questa asimmetria,  il clustering con 4 label non e' sensibile allo shift. 
The results of the clustering are displayed in Figure~\ref{fig:clustering}. The image is shown with reversed intensity contrast to  help the reader to identify the color contours. Each colour contour contains the pixels where the profiles are more similar to the average profile of the same color. The bright centre is described by only one typical profile (azure). This typical profile does not exhibit a strong blue-shift as retrieved in the velocity map because it is an average on many different profiles.
The pixels belonging to the centre of the SGS exhibit a significant asymmetry of the K$_{2 \textrm{r}}$ feature. The strongly enhanced K$_{2 \textrm{r}}$ peak and the blue-shift are the cause of the bright fibrils observed in the nominal line centre of \CaIIK. The clustering results are in agreement with K$_{2 \textrm{r}}$ intensity map. 
The rest of the pixels are labeled with typical QS profiles (green and black contour) or complete absorption profiles (red).
Complete absorption profiles are only observed in the quiet Sun portion of the FOV. %The active region has larger velocity and temperature gradients that prevent the K$_{1_{\textrm{r}}}$ and K$_{1_{\textrm{v}}}$ from disappearing.

%\begin{figure}
%\includegraphics[scale=1]{fig/fig_profiles.eps}\caption{Intensity profiles of \CaIIK\ (left) and \CaII~854.2~nm at the position indicated inf Figure~\ref{fig:linecores}}\label{fig:profiles}
%\end{figure}
%
%\begin{figure}
%\includegraphics[scale=1]{fig/fig_labels_deriv.eps}\caption{\CaIIK line core intensity map. The yellow mask highlight the location of complete absorption profiles.}\label{fig:labels_deriv}
%\end{figure}
%
%\subsection{Power spectrum}
%
%
%
%\begin{figure}
%\includegraphics[scale=1]{fig/pow_b.eps}\caption{Total magnetic field power spectrum calculated for a square image of the annular network (yellow linear fit) and quiet Sun inside the SGS (green linear fit).}\label{fig:pow}
%\end{figure}

\subsection{Evolution}\label{sec:evolution}
\begin{figure}
\includegraphics[scale=1]{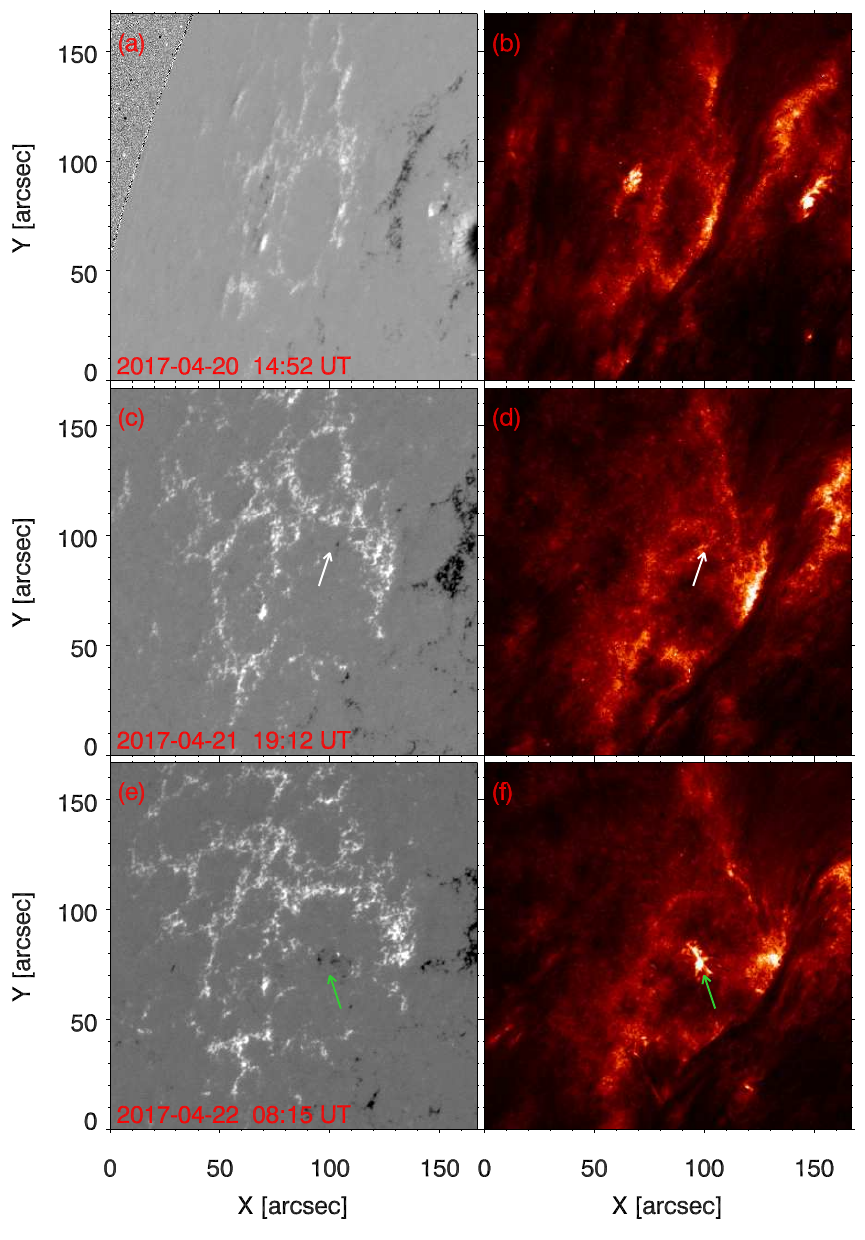}\caption{\textit{Left column:} HMI line-of-sight magnetogram of the super granular structure. \textit{Right column:} AIA~304 images for the same field-of-view as the left column. The first row shows data acquired on the same day of CRISP observations (a-b). The middle and the last row shows data obtained one (c-d) and two (e-f) days later. The green and white arrows indicate flux emergence events. The north direction coincides with the top side of the panels.}\label{fig:fe_sdo}
\end{figure}

\grass{SST observations do not last long enough to catch the long-term evolution of the SGS. To detect any remarkable activity connected with the SGS in the following days, we used co-observations with SDO.}
The left column of Figure~\ref{fig:fe_sdo} shows the time evolution of the SGS magnetogram. We can see that the day after the observation (panel c), partly due to projection effects, more positive network is visible inside the SGS and two days after (panel e), we can eventually distinguish some magnetic cells converging in its centre.

\secgrass{A small flux emergence appears in the north (see the white arrows), but, according to the co-temporal image of AIA~304 channel (panel d), no significant chromospheric effect is associated with it.
Two days after (panel e), we observe a more extended flux emergence in the middle of the SGS (green arrow). 
The corresponding AIA~304 image (panel f) shows a coherently bright structure lasting for almost 30 minutes. }

\grass{Overall, the time evolution (Section~\ref{sec:evolution}) and the velocity distribution (Section~\ref{sec:los}) indicate the centre of the SGS as a favourable location for dynamic events. }

%******************************************************************************************
\section{Discussion \& Conclusions}
%******************************************************************************************
%Although dealing with a large heliocentric angle and a QS low polarimetric signal, 
\grass{We retrieved the local magnetic field in the photosphere and chromosphere, applying respectively ME inversion techniques and the weak field approximation.}

Using the \FeI~6301-6302~\AA\ longitudinal magnetogram, we estimated the flux at the photosphere. The field within the SGS is close to polarity balance with an excess of positive polarity, while the network flux is significantly unbalanced. 

Even though the annular network belongs to a rather active area, the results are in agreement with the flux disequilibrium found in QS \citepads[e.g.][]{2014ApJ...797...49G}.
The largest fraction of magnetic field in the annular network is positive and it is not balanced by negative field at the boundaries nor inside the SGS. \grass{Part of the flux might follow the vertical coronal plumes rooted on the annular network and possibly connect with negative polarity outside the FOV.}

\begin{figure}
\includegraphics[width=8.8cm]{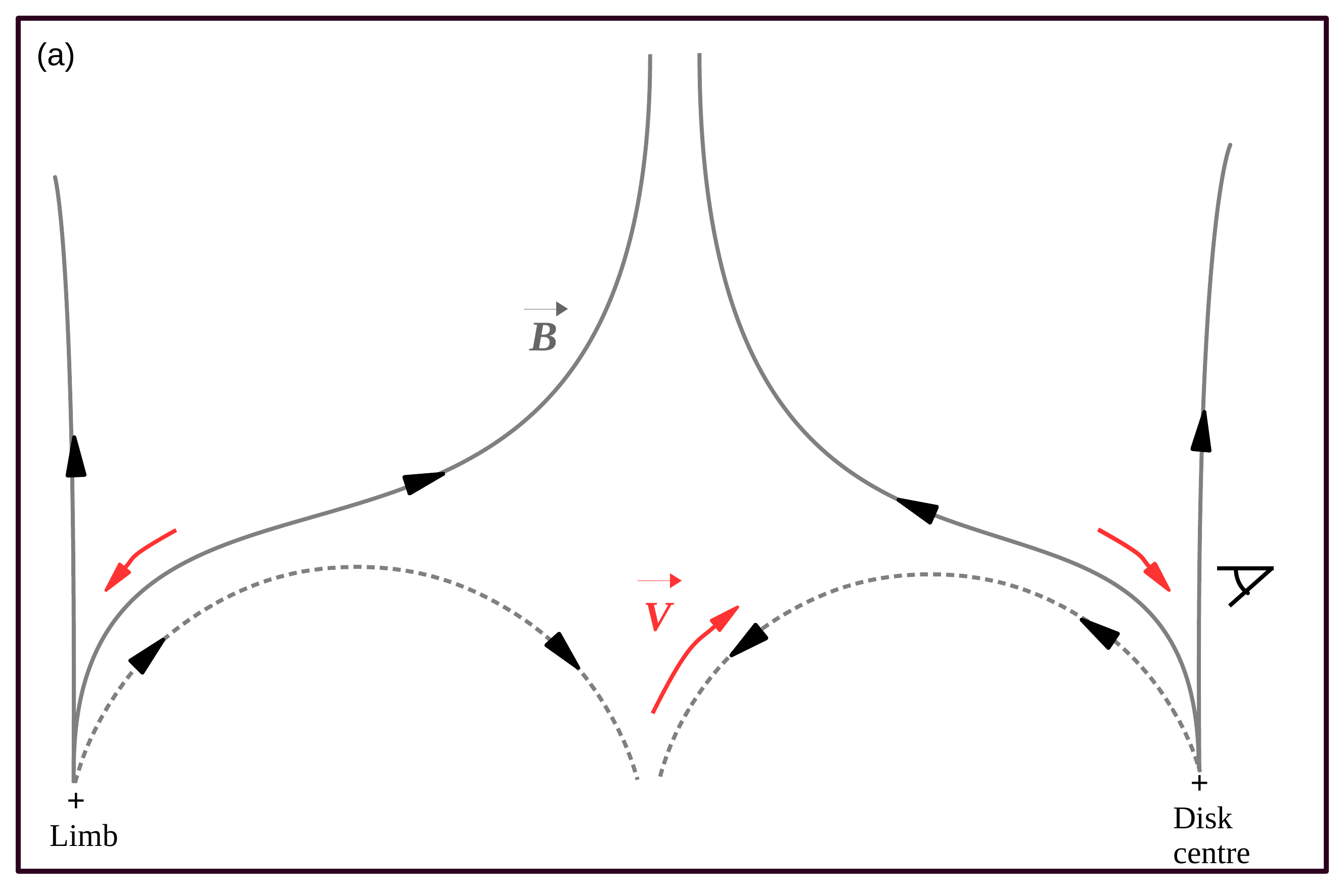}
\includegraphics[width=8.8cm]{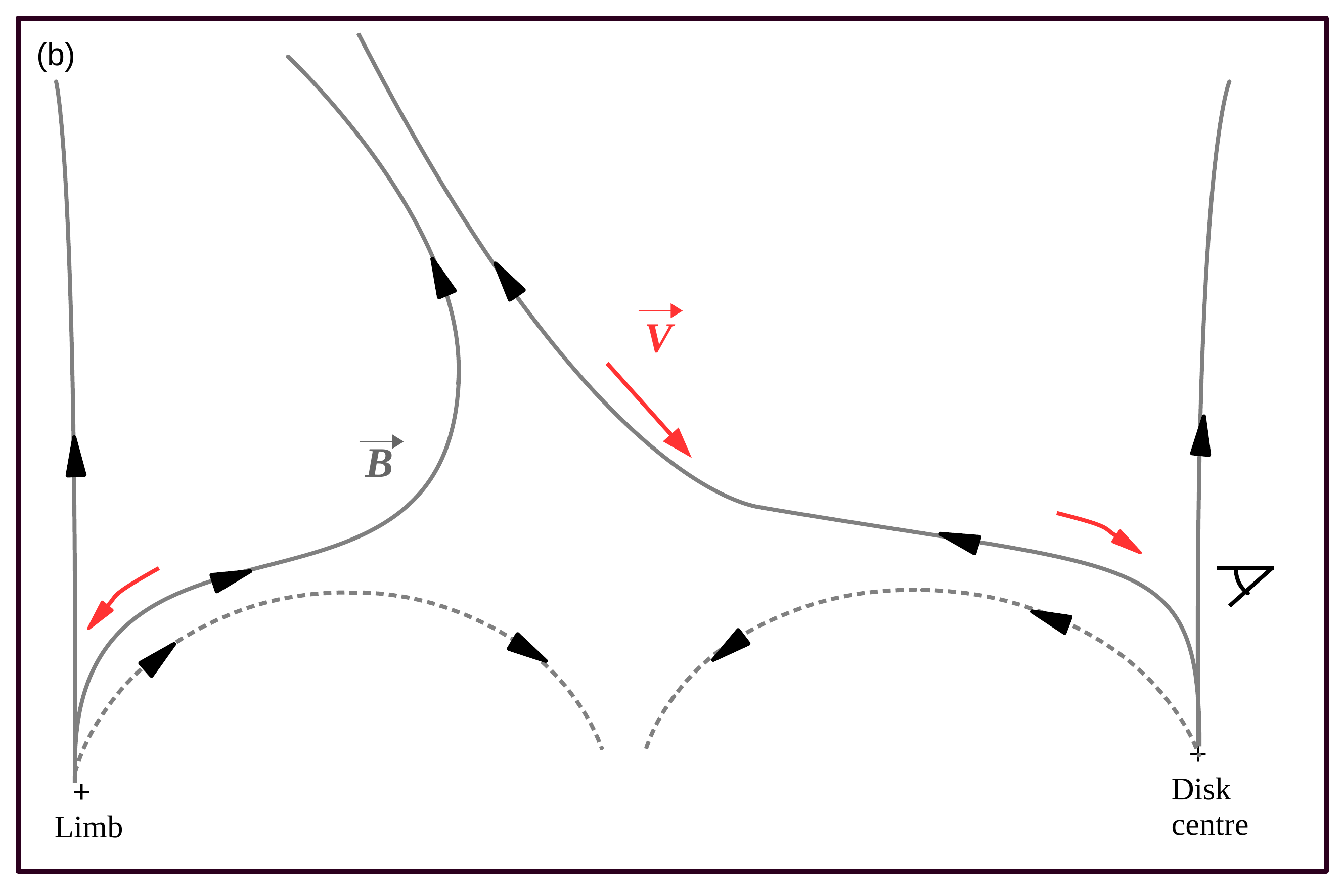}
\caption{Chromospheric magnetic field and velocity models above a section of the supergranular structure. The grey lines indicate the magnetic field lines and the red ones the velocity along the fibrils. Plus signs represent the positive field at the network boundaries. Panel a refers to scenario 1 and panel b to scenario 2 (see the text).}\label{fig:cartoon}
\end{figure}

\grass{Figures~\ref{fig:magneto}f and \ref{fig:incl}a show a rather horizontal chromospheric magnetic field above the SGS. This field is on average directed towards the centre of the structure (Fig.~\ref{fig:magneto}f and Fig.~\ref{fig:incl}b). This configuration is in agreement with the idea of a low-lying canopy field, also suggested by chromospheric images (Fig.~\ref{fig:linecores}) and 171 AIA image (Fig.~\ref{fig:sdo}b).
Limited by the vertical magnetic field map (Fig.~\ref{fig:magneto}f), we cannot clearly see whether the canopy bends upward. However,} if all the magnetic field lines in the centre of the SGS were negative and bent downwards, there would be a significant concentration of negative field in the photosphere.
Since we do not detect it, \grass{and the average field is positively inclined, we suggest that a large part of the chromospheric field lines points upwards, while the remaining returns to the surface, as displayed in Figure~\ref{fig:cartoon}.} These models represents a section of the SGS along the direction disk-centre to limb. The boundaries are indicated by the plus sign. Above the annular network, the field lines are rather vertical, while they bend and form a sort of canopy above the SGS. The models of Figure~\ref{fig:cartoon} show the large-scale behaviour of the magnetic field based  on the popular, but not entirely accepted, magnetic canopy model. \grass{The magnetic field vector that we retrieved from \CaII~8542~\AA\ confirms the inclination measured in the quiet chromosphere by \citetads{2009ASPC..415..327L} and are in agreement with the supergranular topology in the model of \citetads{2009SSRv..144..317W}.
However, the actual \secgrass{sub-canopy magnetic field configuration} is probably more intricate than the models in Fig.~\ref{fig:cartoon}. The low-lying fibril canopy, for instance,} is indicative of a more complex topology, where the QS is strongly connected to the annular network, as proposed by \citetads{2003ApJ...597L.165S}  and suggested also for the chain configuration by \citetads{2011ApJ...742..119R}. 

The fibril arrangement in the SGS is reminiscent of a large rosette structure. Despite being morphologically similar, they form at different locations. The fibrils convergence point of rosettes is usually observed at the vertices of supergranules while the SGS core lies in the centre of the supergranular cell. As a consequence, the rosette core coincides with a strong concentration of magnetic field, which is not observed in this SGS. Both structures have a bright centre, although caused by different reasons: rosette brightening is due to a reduced opacity produced by the magnetic field, while in the SGS is due to a velocity gradient that, in the \CaIIK\ spectral line, blue-shifts the K$_3$ feature and enhances the K$_{2 \textrm{r}}$ peak.

%Figure~\ref{fig:fe_sdo}-e shows that the flux emergence event is located at the vertices of several apparent inner network granules. The magnetic field in this region is of the order 1~kG but there are no available observations of the fibrilar structure for that day to establish whether there is an actual rosette structure. 
%Neither CRISP nor Hinode/SP magnetorgram for the 2017 April 20 show a flux concentration in the middle of the SGS and we do not even observe an inner network granulation. We observe bright fibril bottoms (dark in the reverse intensity of Figure~\ref{fig:zoomin}) produced by a strong velocity gradient along the LOS, which blue-shifts the \CaIIK\ nominal line centre and enhances the K$_{2 \textrm{r}}$ peak.

The chromospheric velocity map (Figure~\ref{fig:vel}b) shows a clear flow towards the network boundaries but the interpretation of the Doppler shift in the SGS core is non-trivial.
To explain the blue-shift observed in the chromospheric lines, especially in the \CaIIK, we propose two scenarios that take into account the magnetic field topology.

\begin{enumerate}

\item In the first scenario, depicted in Figure~\ref{fig:cartoon}a, we assume that the fibrils coinciding with the blue-shifts are totally vertical or slightly inclined towards the observer and, hence, the blue-shift is an actual upflow. Thus, the flow would be directed from the SGS centre towards the network. The SGS is not only the apparent convergence point for the fibrils, but also the location of flux emergence two days after the SST observation. We could therefore interpret the strong blue-shift as a hint of an exiguous undetected flux emergence, taking place during the CRISP observations on 2017 April 20. 
\secgrass{However, we see neither anomalies in the photospheric LOS velocity nor a distortion of the granules as expected in case of flux emergence \citepads{2007A&A...467..703C}. }
%A possible explanation might be that projection effects allow us to see only horizontal motions in the photosphere, while, in the chromosphere, we observe motions channelled into fibrils, at different inclinations. As far it concerns the granulation, the flux emergence has to be so exiguous that it does not visibly alter the intensity nor the shape of the granules.

\item In the second scenario (Figure~\ref{fig:cartoon}b), we interpret the blue-shift as a downflow.
We exclude the possibility of coronal material sliding down along a positive magnetic field canopy, since \halpha\ does not show any trace of coronal rain. So, the downflow has to take place along fibrils inclined towards the limb. The different appearance of the fibrils on the east and west sides of the SGS could be a hint of such a magnetic field asymmetry. 
In that case, we would observe a draining inside loop-shaped fibrils, inclined according to the asymmetric canopy field. 
The draining flow is in agreement with the predominant bulk motion measured by \citetads{1993A&A...271..574T} in a typical rosette, although they also observe an upflow in its core. We believe that this second scenario is more likely since there is no clear indication of flux emergence in the photosphere. 

\end{enumerate}

%******************************************************************************************
\section{Summary}
%******************************************************************************************

We have retrieved the magnetic field and the LOS velocity field in the photosphere and chromosphere of a unipolar region of supregranular size. \grass{We summarise our conclusions as follows:}

\begin{itemize}
\item[•] \grass{The chromospheric magnetic field lines bend above the SGS suggesting a canopy-shaped configuration.}
\item[•] \grass{The photospheric magnetic flux of the entire SGS is unbalanced.}
\item[•] \grass{A strong velocity gradient along the LOS might be responsible for the bright SGS core.}
\item[•] \grass{We interpret the K$_3$ blue-shift observed along the fibrils in the centre of the SGS as a draining flow.}

\end{itemize}

Future studies should address the topology and velocity issues by observing SGSs on multiple days to avoid de-projection mistakes and uncertainties. \grass{Deep spectropolarimetry should be used to retrieve a more accurate inclination of the magnetic field in the chromosphere above SGS. Eventually, studying the time evolution will establish how long the canopy configuration remains coherent.}

\begin{acknowledgements}
The Swedish 1-m Solar Telescope is operated by the Institute for Solar
Physics of Stockholm University in the Spanish Observatorio del Roque
de los Muchachos of the Instituto de Astrof\'{\i}sica de Canarias. 
We made use of resources provided by the Swedish National Infrastructure for Computing (SNIC) at the High Performance Computing Center North at Ume{\aa} University.
JdlCR is supported by grants from the Swedish Research Council (2015-03994), the Swedish National Space Board (128/15) and the Swedish Civil Contingencies Agency (MSB). This project has received funding from the European Research Council (ERC) under the European Union's Horizon 2020 research and innovation programme (SUNMAG, grant agreement 759548).This research was supported by the CHROMOBS and CHROMATIC grants of the Knut och Alice Wallenberg foundation.
This research has made use of NASA's Astrophysical Data System.
\end{acknowledgements}

%%----------------------------------------------------------------------------------------
%%	REFERENCE LIST
%%---------------------------------------------------------------------------------------

%
\bibliographystyle{aa}
%\bibliography{biblio}
%
%

\end{document}